\pdfoutput=1
\documentclass[12pt]{article}

\usepackage{fullpage}
\usepackage{latexsym}
\usepackage{color}
\usepackage{verbatim}
\usepackage{hyperref}
\newcommand{\be}{\begin{equation}}
\newcommand{\ee}{\end{equation}}
\newcommand{\ba}{\begin{eqnarray}}
\newcommand{\ea}{\end{eqnarray}}
\newcommand{\bd}{\begin{displaymath}}
\newcommand{\ed}{\end{displaymath}}

\def\thalf{{\textstyle{\frac{1}{2}}}}

\newcommand{\addchanges}[1]{{\color{black} #1}}
\newcommand{\removechanges}[1]{}%{\color{blue} #1}}

\begin{document}
\title{
{\bf Potentials for Soft-Wall AdS/QCD}}

\removechanges{
\author{J. I. Kapusta}
\affiliation{School of Physics and Astronomy, University of Minnesota, Minneapolis, Minnesota 55455, USA}

\author{T. Springer}
\affiliation{School of Physics and Astronomy, University of Minnesota, Minneapolis, Minnesota 55455, USA}
\affiliation{Department of Physics, McGill University, Montreal, Quebec, H3A 2T8, Canada}
}

\author{{J. I. Kapusta$^a$ and T. Springer$^{a,b}$}
\vspace*{0.1in}\\
$^a${\it School of Physics and Astronomy, University of Minnesota}\\
{\it Minneapolis, Minnesota 55455, USA}\\
$^b${\it Department of Physics, McGill University}\\ 
{\it 3600 University Street, Montreal, Quebec, Canada H3A 2T8}}

\date{April 15, 2010}

%\renewcommand{\baselinestretch}{1.5} %wide line spacing
%\parindent=20pt
%\begin{document}
\maketitle

\begin{abstract}
Soft-wall models in AdS/QCD generally have dilaton and scalar fields
that vary with the fifth-dimension coordinate.  These fields can be
parameterized to yield hadron mass spectra with linear radial
trajectories and to incorporate spontaneous breaking of chiral
symmetry.  We show how to construct scalar potentials which lead to
such solutions.
\end{abstract}

%\maketitle
\newpage
\section{Introduction}
The correspondence between certain gauge theories in four dimensions
and gravitational theories in higher dimensions has proven to be a
very fruitful one.  It allows for the calculation of physical
observables in a strongly coupled gauge theory, where perturbation
theory is inapplicable, by performing a corresponding calculation in a
classical gravitational theory.  This general gauge/gravity duality
was originally inspired by the anti-de Sitter / conformal
field theory (AdS/CFT) correspondence, and as such the metric usually takes the
form of anti-de Sitter space in five dimensions
\cite{Maldacena:1997re,Witten:1998qj,Gubser:1998bc,Klebanov:1999tb}.
QCD, the gauge theory of the strong
interactions, is strongly coupled at low energies.  Heretofore, if one
wanted to understand the structure of hadrons, then the only reliable
means was via numerical computations in lattice gauge theory.  Aside
from lattice gauge theory, one must resort to models.  The discovery
of a gravitational dual to QCD would be a fantastic achievement.
Unfortunately, the gravitational dual is not known, nor is it known
whether one exists even in principle.  Deriving such a duality from a
fundamental theory, such as string theory, is referred to as the top
down approach.  The bottom up approach assumes that such a dual
exists, and it models QCD by an effective five-dimensional gravity
theory.  One would like to incorporate the essential features of QCD
into such a model.  Such effective five-dimensional models are
generically referred to as AdS/QCD, and allow for the computation of
physical quantities in QCD such as mass spectra
\cite{deTeramond:2005su, Erlich:2005qh, Karch:2006pv,
Brodsky:2006uqa, deTeramond:2008ht, Gursoy:2007cb, Gursoy:2007er}, form
factors
\cite{Kwee:2007nq, Grigoryan:2007vg, Grigoryan:2007wn, Abidin:2008ku,
Abidin:2008hn}, and thermodynamic observables \cite{Herzog:2006ra,
BallonBayona:2007vp, Kajantie:2006hv, Gubser:2008ny,Gubser:2008yx,
Andreev:2007zv, Gursoy:2008bu, Gursoy:2008za}

The first model to be constructed via the bottom up approach is
referred to as the hard-wall model \cite{Erlich:2005qh,
Polchinski:2001tt,DaRold:2005zs}; it simply places an infrared cutoff
on the fifth-dimensional coordinate.  This cutoff breaks the conformal
symmetry by hand, and allows for the introduction of the QCD scale.
The soft-wall model improves upon the hard wall, both physically and
mathematically, by using a dilaton field to suppress the infrared
contributions in a softer and continuous fashion \cite{Karch:2006pv}.
\addchanges{
This model is a mathematical improvement on the hard wall, because the
geometry is everywhere continuous and thus avoids discontinuities
and/or singularities which may exist in a hard-wall setup.  The soft
wall model is a physical improvement because
}
the dilaton field is parameterized as a function of the fifth
dimensional coordinate in order to provide linear radial trajectories
for the meson masses.  There have been numerous improvements and
variations on these models, some of which attempt to incorporate
confinement \cite{dePaula:2008fp} and chiral symmetry breaking
\cite{Gherghetta:2009ac}.  In the latter case, a scalar field which is
dual to the condensation of the quark bilinear operator in QCD is
included.  As such, the behavior of this scalar field in the
extra-dimensional geometry corresponds to the properties of chiral
symmetry in the gauge theory.  In an AdS/QCD model, this scalar field
can be parameterized to incorporate both spontaneous and explicit
breaking of chiral symmetry.

As mentioned above, in bottom up models the dilaton and scalar
fields are parameterized to reproduce certain important features of
QCD, namely, confinement and spontaneous and explicit chiral symmetry
breaking.  In almost all cases, these background fields are imposed
by hand, and are not derived as the solution to any equations of
motion.  There are (at least) two reasons that a well-defined action
is desirable within AdS/QCD models.  First, one needs a proper set of
background equations in order to add perturbations to the geometry
which can give access to transport coefficients in the dual field
theory.  Second, a well-defined action may provide insight as to how
such a model could arise from a top down approach.  In order to
meet these goals, one must know the potential which gives rise to such
backgrounds.  Some recent work on this subject was done in \cite{Alanen:2009xs}
where the authors explore the correspondence between the potential,
and the running coupling in the dual field theory.

A notable case where the potential was determined is the dynamical
soft-wall model of \cite{Batell:2008zm}.  There, an auxiliary scalar
field is introduced in addition to the dilaton, and a scalar potential
is derived which has the soft-wall model as a solution to the
equations of motion.  Throughout the aforementioned work, some attempt
is made to identify this auxiliary scalar field with a closed string
tachyon field in string theory.  Instead of this
identification, one can promote the scalar field to a matrix valued
field and identify it as 
\removechanges{the field which is} 
dual to the quark bilinear
operator $\bar{q}q$ which is responsible for chiral symmetry
breaking.  
\removechanges{
If one makes this latter identification, the model of
\cite{Batell:2008zm} breaks chiral symmetry only explicitly; there is no
chiral condensate and hence no spontaneous breaking of the symmetry.
The majority of other AdS/QCD models (with the exception of \cite{Gherghetta:2009ac}) 
have the chiral condensate and quark mass coupled
together, meaning that spontaneous and explicit chiral symmetry
breaking are also coupled (in contrast to QCD).
}
\addchanges{
However, \cite{Batell:2008zm} and the majority of other AdS/QCD models,
such as those in \cite{Karch:2006pv,Kwee:2007nq,Colangelo:2008us}, have
the (light) quark condensate proportional to the (light) quark mass,
$\Sigma \propto m_q$.  This is in contrast to QCD, where the condensate has
a nonvanishing limiting value as $m_q \rightarrow 0$.  An exception is
the phenomenological model of \cite{Gherghetta:2009ac} which parameterizes
the fields in such a way as to allow $m_q$ and $\Sigma$ to be varied
independently.
}

In this paper we attempt to bridge the gap between the dynamical model
\cite{Batell:2008zm} and phenomenological models of chiral symmetry
breaking, such as \cite{Gherghetta:2009ac}, by demonstrating how to
compute a potential given a phenomenological parameterization of the
dilaton and/or scalar fields.  This methodology is an improvement upon
\cite{Batell:2008zm} in two respects.  First, it is more general in
that the potential can be found for many reasonable parameterizations
of the dilaton and scalar fields.  Second, even in the case considered
in \cite{Batell:2008zm}, our resulting potential is both simpler and
less constrained, as discussed in Sec. \ref{powerlaw_section}.

Our paper is organized as follows.  In Sec. \ref{ingredientsection} we
identify the basic ingredients of the AdS/QCD model and elements of
the AdS/CFT dictionary which we will use.  In
Sec. \ref{backgroundequations_section} we give the relevant Einstein
equations, and equations for the background fields which generate the
desired metric.  In Sec. \ref{powerlaw_section} we find the scalar
potential when the fields have power-law profiles.  In
Sec. \ref{chiralsymmetry_section} we show how to find the potentials
for models wherein the fields are allowed to have more complicated
profiles. In Sec. \ref{example_section} we provide concrete examples
of some parameterizations which lead to analytic potentials.  In
Sec. \ref{quadraticchi_section} we discuss what happens if the
potential is restricted to a form which has only quadratic and quartic
terms in the scalar field.  We conclude the paper in
Sec. \ref{conclusion_section}.

\section{Ingredients of the Model}
\label{ingredientsection}

We are interested in dynamically generating phenomenological AdS/QCD
backgrounds.  Following \cite{Batell:2008zm}, we assume that the
matter which supports the metric is a set of two scalar fields: $\phi$
and $\chi$.  These scalar fields interact through a scalar potential
$V(\phi,\chi)$; it will be our goal to determine suitable potentials
which lead to phenomenologically desirable backgrounds.

We would like to have a background which has the following characteristics:
\begin{enumerate}
\item 
The metric in the string frame should be exactly five-dimensional anti-de Sitter space.  
\be
ds^2_{\rm string} = \frac{L^2}{z^2} \left[-dt^2 + dx_i dx^i + dz^2 \right]  
\label{stringmetric}
\ee
Here $L$ denotes the AdS curvature radius, the index $i$ runs over the
three spatial dimensions, and $z$ denotes the extra (fifth) dimension.
\label{item1}
\item
At asymptotically large $z$ the dilaton should behave as $\phi(z) \sim z^2$.
\label{item2}
\item 
  The mass of the dilaton should take a value such that the dual
  operator in the 4D field theory has a physically relevant mass
  dimension.
\label{item6}
\end{enumerate}
Points \ref{item1} and \ref{item2} are necessary ingredients for a
soft-wall model.  Such a background leads to linear radial
trajectories in the resulting meson spectrum ($m_n^2 \sim n$), in
agreement with the data \cite{Karch:2006pv}.  Point \ref{item6} is
related to the mass of the dilaton.  The relationship between the
field's mass and the dimension $\Delta$ of the dual field theory
operator is given by the AdS/CFT dictionary
\cite{Maldacena:1997re,Witten:1998qj,Gubser:1998bc,Klebanov:1999tb}.
\be
\label{massdeltarelation}
m^2 L^2 = \Delta(\Delta - 4) 
\ee
We will show that we are able to keep this parameter arbitrary in many
of our solutions, but of course we want the corresponding gauge theory
operator dimension to be physically relevant, and the mass of the
dilaton should not violate the Breitenlohner-Freedman stability bound
\cite{Breitenlohner:1982bm} which states that $m_\phi^2 L^2 \geq -4$.

The above points are the only necessary requirements for a soft wall
type of model.  If, in addition to this, we would like to associate
the scalar field $\chi$ with the operator responsible for chiral
symmetry breaking, $\bar{q}q$, we require the following:
\begin{enumerate}
\setcounter{enumi}{3}
\item
The mass of the scalar field should be $m_\chi^2 = -3/L^2$.
\label{item3}
\item
At small $z$ the scalar field should behave as
$\chi(z) = A z + B z^3 + \cdot\cdot\cdot$.  
\label{item4}
\item
At asymptotically large $z$ the scalar field should behave as $\chi(z) \sim z$.
\label{item5}
\end{enumerate}
Because the dimension of the operator $\bar{q}q$ is 3, we require that
$m_\chi^2 = -3/L^2$.  If the scalar field $\chi$ is dual to the
operator responsible for chiral symmetry breaking, then in the UV ($z
\to 0$) regime it should have a profile which can be expanded as in
item \ref{item4} above with the coefficients $A$ and $B$ proportional
to the quark mass and the chiral condensate, respectively
(cf. \cite{Klebanov:1999tb,Erlich:2005qh,Gherghetta:2009ac}). Often in
this work we will consider the chiral limit of zero quark mass; in
this case the coefficient $A$ vanishes, and chiral symmetry is broken
spontaneously.  It was also argued in
\cite{Casero:2007ae,Bergman:2007pm,Shifman:2007xn} that the scalar
field $\chi$ should behave linearly at large $z$ so as to realize the
nonrestoration of chiral symmetry at large $n$, again in agreement
with the linear radial trajectories present in the data\footnote
{
  It is important to note there is not a consensus in the literature
  regarding chiral symmetry restoration in highly excited hadrons.  For
  an opposing viewpoint, see \cite{Glozman:2007ek}.  The requirement of
  chiral symmetry restoration at large $n$ would require different
  field profiles at large $z$. We will not address such a model
  within the context of this work.
}.
Note that to properly be dual to this operator, the field $\chi$
should be charged under the bulk chiral symmetry, meaning it should be
complex and matrix valued.  We have introduced a single real scalar
field for simplicity, but we will discuss promotion of this field to a
complex, matrix valued one in Sec. \ref{quadraticchi_section}.

\addchanges{
Before proceeding, let us make some comments on the relation of
our approach to that of standard AdS/CFT.  In the usual
AdS/CFT approach, a conformal field theory (e.g. $\mathcal{N} = 4$ Supersymmetric Yang-Mills theory)
is deformed by introducing operators into the field theory Lagrangian.  
From the gravity point of view, one takes an existing bulk geometry (e.g. $AdS_5$) and
introduces fields \emph{into} this background.  Near the UV boundary, 
these fields act as free, noninteracting fields on $AdS_5$.  As one moves into the
interior of the $AdS$ space, the geometry will no longer be $AdS_5$ due to the backreaction
induced by the new bulk fields.  

In this work, we compute all backreaction by introducing the fields into the
Lagrangian and then solving the equations of motion for all values of
the radial coordinate, including the interior of the $AdS$ space.  As mentioned
in point \ref{item1} above, we have \emph{chosen} the metric to be exactly $AdS_5$, and
thus the potentials which we detail in this work lead to exactly $AdS_5$ metrics
after all backreaction of the fields is taken into account.

Our choice of metric is made simply to reduce the amount of freedom
which exists in the equations of motion; the methods which we detail here could
easily be extended to  metrics which are warped versions of $AdS_5$.  Of course,
one could introduce deformations (in the usual AdS/CFT way) into the backgrounds which we detail below
; such deformations would, in general, introduce backreaction
which would cause the metric to be no longer exactly $AdS_5$ for all values
of the radial coordinate.  

Now that we have detailed the set of requirements for our model and
clarified our approach,
}
%This completes the set of requirements of our model; 
let us now proceed to derive the relevant background equations and
attempt to find potentials and solutions which have the
characteristics we have outlined above.  We will see that it is quite
difficult to create a simple phenomenological model which satisfies
all of the above requirements.

\section{Background Equations}
\label{backgroundequations_section}

In the Einstein frame, the action can be written in its canonical form
\be
\mathcal{S}_E = \frac{1}{16 \pi G_5} \int \, d^{5}x \sqrt{-g_E} 
\left( R_E  - \thalf \partial_\mu \phi \partial^\mu \phi - \thalf \partial_\mu \chi \partial^\mu \chi - V(\phi,\chi) \right) \, . 
\ee
Here $R$ denotes the Ricci scalar, $G_5$ stands for the five-dimensional gravitational constant, and the subscript $E$ denotes the
Einstein frame.  Note that our conventions are such that the fields $\phi$ and 
$\chi$ are dimensionless while $V(\phi,\chi)$ has dimensions of energy squared.  The energy momentum tensor
which is derived from this action is
\ba
8 \pi G_5 T_{\mu \nu} 
&=& \thalf \left(\partial_\mu \phi \partial_\nu \phi + \partial_\mu \chi \partial_\nu \chi - g_{\mu \nu} \mathcal{L} \right) \, , 
\label{scalarTuv}
\\
\mathcal{L} &\equiv& \thalf \partial_\lambda \phi \partial^\lambda \phi
+ \thalf \partial_\lambda \chi \partial^\lambda \chi + V(\phi,\chi) \, .
\ea
We assume that in the string frame there is a nontrivial coupling of
the dilaton to the Ricci scalar.  As such, the string and Einstein
frame metrics are related by the conformal transformation
\be
\label{einsteinmetric}
g_{\mu \nu}^{\rm string} = e^{2 a \phi} g_{\mu \nu}^{E} \, , 
\ee
where $a$ is a constant which depends on the coupling of the dilaton
to the Ricci scalar in the string frame action.  Because the metric is
static and depends only on the extra-dimensional coordinate $z$, we
make the usual assumption that the fields themselves are only
functions of this coordinate.

There are four nontrivial background equations.  We will work with
the following combinations, where $G_{\mu \nu}$ denotes the Einstein
tensor:
\ba
g^{tt}G_{tt} - g^{zz}G_{zz} &=& 8 \pi G_5 \left(g^{tt}T_{tt} - g^{zz}T_{zz}\right) = \frac{1}{2}g^{zz} \left( [\phi'(z)]^2 + [\chi'(z)]^2 \right) \, ,\\
g^{tt}G_{tt} + g^{zz}G_{zz} &=& 8 \pi G_5 \left(g^{tt}T_{tt} + g^{zz}T_{zz} \right) = -V(\phi,\chi) \, ,\\
\Box \phi &=& \frac{\partial V}{\partial \phi} \, ,\\
\Box \chi &=& \frac{\partial V}{\partial \chi} \, .
\ea
As usual, $\Box \equiv \nabla_\mu \nabla^\mu$ and $\nabla_\mu$ denotes
the covariant derivative with respect to the background metric.  Using
the presumed form of the metric expressed by Eqs. (\ref{stringmetric})
and (\ref{einsteinmetric}), these become
\ba
6 a \phi''(z) + [\phi'(z)]^2 \left(6a^2 -1 \right) - [\chi'(z)]^2 + \frac{12 a \phi'(z)}{z} &=& 0 \, ,
\label{EE2}\\
3 e^{2 a \phi(z)} \frac{z^2}{L^2} \left[ a \phi''(z) 
- 3 a^2 [\phi'(z)]^2 - \frac{6}{z} a \phi'(z) - \frac{4}{z^2}\right] &=& V(\phi(z),\chi(z)) \, ,
\label{EE3}\\
e^{2 a \phi(z)} \frac{z^2}{L^2} \left[\phi''(z) - 3 a [\phi'(z)]^2 - \frac{3 \phi'(z)}{z} \right] 
&=& \left. \frac{\partial V}{\partial\phi} \right|_{\phi = \phi(z), \chi = \chi(z)} \, ,
\label{EE4}\\
e^{2 a \phi(z)} \frac{z^2}{L^2} \left[\chi''(z) - 3 \chi'(z)\left(a \phi'(z) + \frac{1}{z} \right)\right] 
&=& \left. \frac{\partial V}{\partial\chi} \right|_{\phi = \phi(z), \chi = \chi(z)} \, . 
\label{EE5}
\ea
These equations are not all independent.  Because the potential
depends on $z$ only through the fields, we have
\be
\frac{d}{dz} V(\phi(z), \chi(z)) = \frac{\partial V}{\partial \phi} \phi'(z) + \frac{\partial V}{\partial \chi} \chi'(z) \, .
\ee
This relation allows one to eliminate either (\ref{EE5}) or (\ref{EE4}).

So far we have kept the constant $a$ arbitrary; its value can be
fixed by examining (\ref{EE2}).  Consider this equation in the limit
of large $z$, where the dilaton is required to be quadratic in $z$.
Assuming
\be
\phi(z) = \lambda z^2 \, ,
\ee
with $\lambda$ a constant, Eq. (\ref{EE2}) becomes
\be
[\chi'(z)]^2 = 36 a \lambda + 4 \lambda^2 z^2 \left(6a^2 -1 \right) \, .  
\label{EE2largez}
\ee
In the limit of large $z$, the constant term can be dropped.  The
solution to the resulting differential equation gives $\chi(z) \sim
z^2$ at large $z$.  This is in contradiction with the desired behavior
detailed in Sec. \ref{ingredientsection}.  The only way to avoid this
problem is by choosing
\be
a = \pm 1 / \sqrt{6}
\ee
so that the $z^2$ term drops out in the above equation.  With this
choice, one is able to have the desired quadratic dilaton and linear
scalar field at large $z$.  We must choose the positive sign so as to
keep $\chi$ real.  Thus, we will fix $a = 1 / \sqrt{6}$ for the
remainder of this work.  
\addchanges{
It is quite interesting that this value of $a$ appears frequently in the literature
and can arise quite naturally from noncritical string theory 
\cite{Gursoy:2007cb, Gursoy:2007er, Batell:2008zm, Alvarez:1999mp, Csaki:2006ji}.
}
With this value of the $a$, the string frame action is
\be
\mathcal{S}_{\rm string} = \frac{1}{16 \pi G_5} \int \, d^{5}x \sqrt{-g_s} 
e^{-2 \Phi} \left( R_s  + 4 \partial_\mu \Phi \partial^\mu \Phi - \thalf \partial_\mu \chi \partial^\mu \chi - e^{-4 \Phi /3} V(\phi,\chi) \right) \, .
\label{stringaction} 
\ee
Here we are using the subscript $s$ to distinguish the string frame
from the Einstein frame.  The field $\Phi$ is a scaled version of
$\phi$,
\be
\Phi = \sqrt{\frac{3}{8}} \phi \, .
\ee
The authors of \cite{Batell:2008zm} considered this action based on
string theory considerations.  In fact, the normalization of the
kinetic term for the dilaton above is exactly the same as that which
often appears in a low energy effective action for string theory (cf. Sec. 3.7 of
\cite{Polchinski:1998}).  We have shown that this is the \emph{only}
possible dilaton-scalar action which is consistent with our desired
behavior outlined in Sec. \ref{ingredientsection}.  In other words, if
the kinetic term in (\ref{stringaction}) were normalized differently,
one could not satisfy all of the requirements in
Sec. \ref{ingredientsection}.

\section{Power-Law Solutions}
\label{powerlaw_section}

Let us first try to construct potentials which have solutions where
$\chi$ is purely a power-law in $z$.  As such, we make the ansatz
\be
\chi(z) = \chi_0 z^n \, .
\label{chipowerlaw}
\ee
Of course, only certain powers of $z$ appear in the ingredients of our
model outlined in Sec. \ref{ingredientsection}, but it is just as easy
to work with a general $n$ at this stage.  Equation (\ref{EE2}) gives
the solution for $\phi$,
\be
\phi(z) = \frac{n \sqrt{6}}{12(1+2n)} \chi_0^2 z^{2n} \, .  
\label{phipowerlaw}
\ee
Here we have assumed a Dirichlet boundary condition on $\phi$ so that
$\phi(0) = 0$.  With this choice, the Einstein frame metric is asymptotically anti-de Sitter.

Inserting the solutions for the fields into (\ref{EE3}) and (\ref{EE4})
yields the system
\be
V(\phi(z),\chi(z)) = \frac{e^{2\phi / \sqrt{6}}}{L^2} \left[-12 + \frac{(\chi_0 n z^n)^2}{2(1+2n)}(2n - 7) - \frac{(\chi_0 n z^n)^4}{4(1+2n)^2} \right] \, , 
\label{EE3powerlaw}
\ee
\be
\left. \frac{\partial V}{\partial\phi} \right|_{\phi = \phi(z), \chi = \chi(z)} = \frac{2e^{2\phi / \sqrt{6}}}{\sqrt{6}} 
\left[\frac{(\chi_0 n z^n)^2}{2(1+2n)}(2n - 4) -  \frac{(\chi_0 n z^n)^4}{4(1+2n)^2} \right] \, .
\label{EE4powerlaw} 
\ee
As mentioned previously, we need not consider Eq. (\ref{EE5})
since it is not independent of the two listed above.

The challenge now is to determine the potential $V(\phi,\chi)$.  The
authors of \cite{Batell:2008zm} assume that the potential can be
derived from a superpotential and use this fact to help determine $V$.
We will take a different route.  By examining the structure of the
above equations we notice two facts.  First, both equations contain
the same exponential factor.  This leads us to believe that the potential
can be written as
\be
V(\phi,\chi) = e^{2 \phi / \sqrt{6}} \tilde{V}(\phi,\chi) \, .  
\ee
This form of the potential is also natural on the grounds that the
exponential factor arises due to the transformation between the string
and Einstein frames.  ($\tilde{V}$ is the string frame potential.)
Second, we notice that the only powers of $z$ which appear in the
above system of equations are $z^{2n}$ and $z^{4n}$.  Motivated by
this, and the fact that we know the power-law behavior of both $\phi$
and $\chi$, we make the ansatz
\be
V(\phi,\chi) = \frac{e^{2\phi / \sqrt{6}}}{L^2} \left[c_0 + c_1\phi + c_2\chi^2 + c_3 \phi^2 + c_4 \chi^4 + c_5 \phi \chi^2 \right] \, .
\label{Vansatz}
\ee
The terms proportional to $\phi$ and $\chi^2$ give rise to the terms
containing $z^{2n}$, while the terms proportional to $\phi^2$,
$\chi^4$, and $\phi \chi^2$ give rise to the $z^{4n}$ terms.  One
should then insert this ansatz into the background equations
(\ref{EE3powerlaw}) and (\ref{EE4powerlaw}), match the coefficients of
$z$ on each side of the equation, and try to solve for the
coefficients $c_0...c_5$.

The resulting system of equations \emph{does} have a solution.  Even
more remarkably, the solution only determines five of the six
coefficients.  For example, $c_1...c_6$ can be written in terms of  $c_3$ as follows:
\ba
c_0 &=& -12 \, , \\
c_1 &=& 4 \sqrt{6} \, , \\
c_2 &=& \frac{n(n-4)}{2}, \\
c_4 &=& \frac{n^2(c_3 - 6n(1+n))}{24(1+2n)^2} \, ,\\
c_5 &=& \frac{n(3n-c_3)}{\sqrt{6}(1+2n)} \, .
\ea

At first sight, one may be concerned about the term linear in $\phi$ in (\ref{Vansatz}) .  But when one expands the Einstein frame
potential, one finds 
\addchanges{
\be
V(\phi, \chi) \approx \frac{-12}{L^2} + \frac{4+c_3}{L^2} \phi^2 + \frac{n(n-4)}{2L^2} \chi^2 
- \left[\frac{n \left(4+c_3+4 n-2 n^2\right)}{\sqrt{6} (1+2 n)L^2}\right] \phi  \chi ^2...
\label{Vsmallfield}
\ee
}
In fact, the value of $c_1$ which solves the background equations is
also the value which ensures the expansion of the potential has no
linear terms.  This form of the potential is exactly what we would
expect from the AdS/CFT correspondence.  The first term is the usual
cosmological constant term, which is followed by the mass terms
of the two scalar fields.  Evidently, we should identify
\be
c_3 = \frac{(m_\phi L)^2 - 8}{2} \, .
\ee
\removechanges{
Usually the dilaton is presumed to be dual to the dimension 4 gluon
condensate operator, and hence has $m_\phi = 0$.  However, in order to
be completely general, we will leave its mass arbitrary whenever possible.}

\addchanges{
Some comments are in order regarding this arbitrary mass.  Astute readers will 
notice that an arbitrary mass term $m_\phi$ and a field $\phi$ which behaves
as $z^{2n}$ near the $AdS$ boundary, do not agree with the standard AdS/CFT
correspondence where the fields near the boundary behave as
\be
	\phi(z \to 0) \sim A z^{\Delta}(1 + ...) + B z^{4-\Delta}(1+...).
\ee
\removechanges{
The reason for this discrepancy can be traced back to the fact that for 
our solution we have two scalar fields with particular scaling $\phi \sim \chi^2$
near the boundary, and due to the presence of the term proportional to $\phi \chi^2$
in (\ref{Vsmallfield}).} 
\addchanges{
This discrepancy is due to two facts.  First, we have two scalar
fields with particular scaling $\phi \sim \chi^2$ near the boundary,
and second there is an interaction term proportional to $\phi \chi^2$
in (\ref{Vsmallfield}).
}
Consider the background
equation (\ref{EE4}) for the field $\phi$ near $z = 0$ where the fields 
are small.  The right hand side depends on $\partial V /\partial \phi$, and using
(\ref{Vsmallfield}), we see that the leading term is $m_\phi \phi$ as expected.  
However, the final term in (\ref{Vsmallfield}) contributes a term
proportional to $\chi^2$.  For the solutions we consider, $\phi$ is proportional to $\chi^2$ near
the boundary, and thus this term cannot be neglected with respect to
the term $m_\phi \phi$ in (\ref{EE4}).  This is significant because it
means that the equation of motion (\ref{EE4}) does \emph{not} reduce
to that of a free scalar field on $AdS_5$ near the boundary.  The
original papers on AdS/CFT (cf. \cite{Klebanov:1999tb}), assumed the
fields in question were free (i.e. noninteracting) near the boundary,
and this is the reason for the discrepancy of our results with the
usual AdS/CFT ones.  \emph{If} one requires agreement with the usual
AdS/CFT prescription, one can easily do so by requiring that the
interaction term in (\ref{Vsmallfield}) vanish.  This leads to the
requirement that
\be
	4+c_3 = 2n^2 - 4n,
\ee
and hence
\be
	(m_\phi L)^2 = 2n(2n-4),
\ee
which is in agreement with standard AdS/CFT.  For a quadratic dilaton
($n=1$), we have $(m_\phi L)^2 = -4$, the same value found in
\cite{Batell:2008zm}.  Unfortunately, this requires an operator
of dimension $\Delta_{\phi} = 2$ which does not correspond to any
local gauge invariant operator in QCD.  For the remainder of this
work, we will choose to continue to keep $m_{\phi}$ arbitrary as much
as possible.  Because the standard AdS/CFT dictionary does not apply
except for particular values of $m_\phi$, field theory interpretations
of the operator dual to $\phi$ will require a careful analysis which
is beyond the scope of this work.  This issue does not arise for the
field $\chi$.  Regardless of the value of $m_\phi$,} 
\removechanges{
It is also interesting to note that} 
\be
m_\chi^2 L^2 = n(n-4) \, .%,
\ee
\removechanges{as this is very reminiscent of (\ref{massdeltarelation}).}
Furthermore, if we require that $m_\chi^2 L^2 = -3$ as outlined in
Sec. \ref{ingredientsection}, the only acceptable values of $n$ are
$n=1$ and $n=3$.  (This fact was also noted in the recent work
\cite{Afonin:2010fr}.)  These are exactly the two different powers of $n$
which appear in the low and high $z$ regions in a desired
phenomenological model.  Putting everything together, we can write the
potential as
\ba
V(\phi,\chi) &=& \frac{e^{2\phi / \sqrt{6}}}{L^2} \biggl\{-12 + 4 \sqrt{6} \phi + \frac{(m_\phi L)^2 - 8}{2} \phi^2 
+ \frac{n(n-4)}{2}\chi^2 \biggr.
\label{powerlawpotential} \\
&+& \biggl. \left[8 - (m_\phi L)^2+6n \right]k \phi \chi^2 
+ \frac{1}{2} \left[(m_\phi L)^2 -8 -12n(1+n)\right]k^2 \chi^4 \biggr\} \, . \nonumber
\ea
Here we have defined
\be
k \equiv \frac{n\sqrt{6}}{12(1+2n)} \, .
\ee
This potential admits a solution where the string frame metric is
given by (\ref{stringmetric}), and the fields $\phi$ and $\chi$ have
power-law profiles (\ref{chipowerlaw}) and (\ref{phipowerlaw}).  For
the case of $n=1$, we have the dynamical soft-wall model with a
quadratic dilaton and linear scalar field.  The potential listed above
is an alternative to that given in \cite{Batell:2008zm}.  We believe
that this potential is an improvement over the latter because it is
simpler and because the mass of the dilaton can be chosen at
will.
\removechanges{\footnote{In \cite{Batell:2008zm}, the mass of the dilaton was
fixed at $m_\phi^2 L^2 = -4$.  The corresponding dual operator in the
$4D$ field theory has mass dimension 2 which does not correspond to
any local gauge invariant operator in QCD. Our model
does not suffer from this difficulty.}}

Of course, the potential listed above is not unique.  One could add
terms provided they vanish upon application of the equations of motion.  If a term
$\Delta V(\phi,\chi)$ is added, the above solution will still be a
solution of the new potential provided that
\ba
\Delta V(\phi(z),\chi(z)) &=& 0 \, , \\ 
\left. \frac{\partial (\Delta V)}{\partial\phi} \right|_{\phi = \phi(z), \chi = \chi(z)} &=& 0 \, , \\
\left. \frac{\partial (\Delta V)}{\partial\chi} \right|_{\phi = \phi(z), \chi = \chi(z)} &=& 0 \, ,
\ea
where $\phi(z)$ and $\chi(z)$ are given in (\ref{chipowerlaw}) and
(\ref{phipowerlaw}).  As an example, a term proportional to
\bd
(\phi - k \chi^2)^l
\ed
will not change the equations of motion provided $l \geq 2$.  

\section{Solutions for General Parameterizations}
\label{chiralsymmetry_section}

We are now in a position to find potentials for more complicated field
profiles, those which are not exactly power laws.  As mentioned previously, at small $z$
we would like either $\chi \sim z$ or $\chi \sim z^3$ depending on
whether the quark mass is zero or not.  At large $z$ we require $\chi
\sim z$; this is necessary for a quadratic dilaton (and hence linear
radial trajectories) by (\ref{EE2}).

By examining Eq. (\ref{powerlawpotential}) for $n = 1$ and $n
= 3$, one notices that the only differences between these two
potentials are the coefficients of the $\chi^4$ term and the $\phi
\chi^2$ term.  Motivated by this, let us make an ansatz that the
potential can be written
\be
	V(\phi,\chi) = \frac{e^{2\phi/\sqrt{6}}}{L^2} \biggl[ -12 + 4\sqrt{6}\, \phi 
	  + \left(\frac{(m_\phi L)^2 - 8 }{2}\right) \phi^2 
	  - \frac{3}{2} \chi^2 \biggr. \nonumber + \biggl. f_1(\chi)  \chi^4 + f_2(\chi) \phi \chi^2 \biggr]. \nonumber
\label{interpolatingpotential}
\ee
The functions $f_1$ and $f_2$ could, in general, also depend on $\phi$,
but we have chosen them to only be functions of $\chi$ for simplicity.
Let us also assume that $\chi$ is a monotonically increasing function
of $z$ such that when $z$ is large, $\chi$ is large, and when $z$ is
small, $\chi$ is small.  Then, to have the desired scalar field
profile, we simply require that
\ba
	f_1(\chi \to \infty) &=&   \frac{1}{432} \left(L^2 m_\phi^2 - 32 \right), \\
	f_2(\chi \to \infty) &=&   \frac{\sqrt{6}}{36} \left(14 - L^2 m_\phi^2\right), 
\ea
and
\ba
	f_1(\chi \to 0) &=& \frac{1}{432}\left(L^2 m_\phi^2 -32 \right), \\
	f_2(\chi \to 0) &=& \frac{\sqrt{6}}{36} \left(14 - L^2 m_\phi^2\right),
\ea
if the quark mass is nonzero.  If the quark mass is zero, we desire
\ba
	f_1(\chi \to 0) &=& \frac{3}{784}\left(L^2 m_\phi^2 -152 \right), \\
	f_2(\chi \to 0) &=& \frac{\sqrt{6}}{28} \left(26 - L^2 m_\phi^2\right) \, ,
\ea
with $f_1$ and $f_2$ smooth functions of
$\chi$.  In other words, when the fields are large, the potential
should take the form of (\ref{powerlawpotential}) with $n=1$, and when
the fields are small, the potential should take the form of
(\ref{powerlawpotential}) with either $n=3$ or $n=1$.  One cannot
choose any convenient functions $f_1$ and $f_2$; they must be
consistent with the equations of motion.  To see this, for the moment
let us assume that the function $\chi(z)$ is known and exhibits the
desired low and high $z$ behavior.  Let us now determine the functions
$f_1(\chi)$ and $f_2(\chi)$.

One must go back to the original background equations (\ref{EE2}) - (\ref{EE5}) using this potential ansatz.  
First, one can solve (\ref{EE2}) for $\phi(z)$ in terms of $\chi(z)$.  The solution is
\be
\phi(z) = \phi_0 + \frac{\phi_1}{z} + \frac{1}{\sqrt{6}}
\int_0^z \frac{dy}{y^2} \int_0^y x^2 \left[\chi'(x)\right]^2 \,dx \, .  
\ee
where $\phi_0$ and $\phi_1$ are integration constants.  We would like $\phi(0) = 0$ so that the Einstein frame metric is asymptotically anti-de Sitter.  In order to satisfy this boundary condition, both integration constants must vanish.  For convenience, we now perform an integration by parts so that 
\be
\phi(z) = \frac{1}{\sqrt{6}}\left( \int_0^z x \left[\chi'(x)\right]^2 \,dx - \frac{1}{z} \int_0^{z}
x^2 \left[\chi'(x)\right]^2\, dx \right) \, .
\ee
Next, one can take a linear combination of (\ref{EE3}) and (\ref{EE4})
such that both the $\phi''$ and $\phi'^2$ terms are eliminated.  This
equation is
\be
\phi'(z) z + \alpha \phi(z) - \frac{\chi^2(z)}{3} f_2(\chi(z)) = 0 \, ,  
\ee
where 
\be
\alpha \equiv \frac{1}{3} \left[ 8 - (m_\phi L)^2 \right] \, .
\ee
Substituting in the solution for $\phi(z)$ here, one can solve for $f_2$:
\be
f_2(\chi(z)) \chi^2(z) = \frac{3}{\sqrt{6}} \left[ \frac{(1-\alpha)}{z} \int_0^z x^2 \left[\chi'(x)\right]^2 \,dx 
+ \alpha \int_0^z x \left[\chi'(x)\right]^2 \,dx \right] \, . 
\ee
Finally, one has to go back to either (\ref{EE3}) or (\ref{EE4}) and solve for $f_1(\chi)$.  First, 
let us define some convenient notation,
\be
I(z) \equiv \int_0^z x \left[\chi'(x)\right]^2\,dx \, . %\, , \\
\ee
%%%%%%%%%%%%%%%%%%%%%%%
%\removechanges{
%\be
%I_2(z) \equiv \frac{1}{z}\int_0^z x^2 \left[\chi'(x)\right]^2 \,dx \, .
%\ee
%}
%%%%%%%%%%%%%%%%%%%%%%%55
Then one finds the following solution for $f_1(\chi)$.
\bd
f_1(\chi(z)) \chi^4(z) = \frac{z^2}{2} \left[\chi'(z)\right]^2 + \frac{3}{2}\chi^2(z) + \frac{1}{6(1-\alpha)} f_2^2(\chi(z)) \chi^4(z)
\ed
\be
- \frac{I(z)}{2} \left[ (m_\phi L)^2 + 3 \alpha + \frac{2}{\sqrt{6}(1-\alpha)} f_2(\chi(z)) \chi^2(z) - \frac{\alpha}{2(1-\alpha)} I(z) \right]
\ee
Now we see how the functions $f_1$ and $f_2$ are correlated with the solution $\chi(z)$.  

Of course we are not finished, because the potential should be a
function of the fields $\phi$, $\chi$ only and should not depend
explicitly on the coordinates.  The expressions for $f_1$ and $f_2$
above need to be rephrased so that they only depend on the field
$\chi$.  To achieve this, instead of parameterizing $\chi(z)$, we should
specify the inverse relationship $z(\chi)$.  As before, we assume that
$z$ is a monotonically increasing function of $\chi$.  For the correct
asymptotic behavior, we require
\be
	z(\chi \to \infty) \sim \chi \, ,
\ee
and
\be
	z(\chi \to 0) \sim \chi \, ,
\ee 
if the quark mass is nonzero, and
\be
	z(\chi \to 0) \sim \chi^{1/3} \, ,
\ee 
if the quark mass is zero.  It is now a simple matter to transform the potential using the relation
\be
\chi'(z) = \frac{1}{z'(\chi)} \, .
\label{zchicalc}
\ee
For example:
\be
I(\chi) = \int_{\chi(0)}^{\chi(z)} z(\chi) \frac{1}{\left[z'(\chi)\right]^2} \frac{dz}{d\chi}\,d\chi = \int_{0}^{\chi} \frac{z(\chi)}{z'(\chi)} \,d\chi \, .
\label{I1chidef}
\ee
This allows us to compute the potential as a function of the
fields only.  For convenience, define
\ba
	\xi_1(\chi) &\equiv& \frac{\alpha}{1-\alpha} \int_0^\chi \frac{z(\chi')}{z'(\chi')} \, d\chi', \label{xi1def}\\
	\xi_2(\chi) &\equiv& \frac{1}{z(\chi)} \int_0^\chi \frac{z(\chi')^2}{z'(\chi')} \, d\chi'. \label{xi2def} 
\ea
Then the solutions for $\phi$ and the potential are
\be
\phi(\chi) = \frac{1}{\sqrt{6}} \left[ \frac{1-\alpha}{\alpha} \xi_1(\chi) - \xi_2(\chi) \right] \, ,
\label{finalphisoln}
\ee
\be
\chi^2 f_2(\chi) = \frac{3}{\sqrt{6}} (1-\alpha) \left[\xi_1(\chi) + \xi_2(\chi) \right] \, ,
\ee
\be
\chi^4 f_1(\chi) = \frac{1}{2} \left(\frac{z(\chi)}{z'(\chi)}\right)^2 + \frac{3}{2} \chi^2 
+ \frac{1-\alpha}{4 \alpha} \Biggl\{(\alpha-1)[\xi_1(\chi) + \xi_2(\chi)]^2 + \xi_2^2(\chi) - 16 \xi_1(\chi) \Biggr\} \, . 
\ee
In all, 
\bd
V(\phi,\chi) = \frac{e^{2\phi / \sqrt{6}}}{L^2} \left\{-12 
+ \frac{\sqrt{6}}{2}\phi\Bigl[8 + (1-\alpha)\left(\xi_1(\chi) + \xi_2(\chi)\right) \Bigr] - \frac{3 \alpha}{2} \phi^2  
\right.
\ed
\be
+ \left. \frac{1}{2}\left(\frac{z(\chi)}{z'(\chi)}\right)^2
+ \frac{\alpha-1}{4 \alpha} \Bigl[(1-\alpha)[\xi_1(\chi) + \xi_2(\chi)]^2 - \xi_2^2(\chi) + 16 \xi_1(\chi) \Bigr]\right\} \, .
\label{finalVsoln}
\ee
This set of equations is one of the central results of this work.
Given a phenomenological parameterization $z(\chi)$, one needs to do
two integrals to determine $\xi_1$ and $\xi_2$, after which point the
potential which gives rise to the desired solution can be determined.
In addition, one can see how the solution for $\phi$ must be correlated
with such a parameterization from (\ref{finalphisoln}).  

The simplicity of the potential is dependent on the simplicity of the
$\xi$ functions, which are in turn related to the parameterization
$z(\chi)$.  We are free to choose $z(\chi)$ at will.  Unfortunately,
there is no guarantee that the potential has an analytic form, as the
integrals (\ref{xi1def}) and (\ref{xi2def}) cannot always be done in
closed form.  In the following section, we will give a few examples of
parameterizations $z(\chi)$ which lead to analytic potentials.

\section{Examples}
\label{example_section}

A simple parameterization $z(\chi)$ which has the correct behavior when the quark mass is zero is
\be
z(\chi) = \gamma \left[\chi + \left(\frac{\chi}{\beta^2} \right)^{1/3} \right] \, ,
\ee
where $\beta$ is a positive dimensionless constant and $\gamma$ is a
positive constant with dimension of length.  Such a parameterization
allows for an analytic potential.  The relevant functions which appear
in the potential are
\be
\xi_1(\chi) = \frac{\alpha}{18 \beta^2(\alpha - 1)} \left[ 6y^{2/3} - 9
y^{4/3} - 9 y^2 - 2 \ln \left(1 + 3 y^{2/3} \right) \right] \, ,
\ee
\be
\xi_2(\chi) = \frac{4}{27 \beta^2 \left(y^{1/3} + y \right)} 
\Biggl[y^{1/3} \Biggr. - \Biggl.y + \frac{27}{15}y^{5/3}+\frac{135}{28}
y^{7/3} + \frac{27}{12}y^3 \Biggr.
- \Biggl. \frac{1}{\sqrt{3}}\arctan \left(\sqrt{3} y^{1/3} \right) \Biggr] \, ,
\ee
where $y \equiv \beta \chi$.  In fact, the above parameterization can
be generalized to
\be
z(\chi) = \gamma \left(\frac{\chi}{\beta^2}\right)^{1/3}
 \left[ 1 + (\beta \chi)^{2/3n} \right]^n \, .
\ee
Both of the integrals involved in the computation of $\xi_1$ and
$\xi_2$ can be done analytically without the use of special functions
if $n$ is an integer (though $n$ must be positive in order for the
field $\chi$ to have the correct asymptotic behavior).  The complexity
of $\xi_1$ and $\xi_2$ appears to increase with increasing $n$, so we
have quoted the simplest example above with $n=1$.

Another parameterization for zero quark mass which leads to the desired behavior is 
\be
z(\chi) = \frac{ \gamma \chi^{1/3}}{\beta^{2/3}\left\{1 + (\beta \chi)^{1/3} \left[ \arctan \left((\beta \chi)^{1/3}\right) - \pi/2 \right]\right\} } \, .
\ee
This parameterization was found by examining the behavior of the
integral appearing in $\xi_2(\chi)$.  The integrand must behave as
$\chi^{4/3}$ for small $\chi$, and as $\chi^2$ for large $\chi$.
Solving the differential equation
\be
\frac{z^2(\chi)}{z'(\chi)} = 3\frac{\gamma}{\beta^2}
\left[ (\beta \chi)^{4/3} + (\beta \chi)^2 \right]
\ee
leads to the above parameterization.  With this parameterization, the solutions for $\xi_1$ and $\xi_2$ are 
\ba
\xi_1(\chi) &=& \frac{\alpha}{7\beta^2(1-\alpha)} \left[-y^{2/3}+ \frac{1}{2} y^{4/3} + \frac{61}{6} y^2 + 7 y^{8/3} \right. \nonumber \\
 &+& \left. \left(7y^3+ 9y^{7/3}\right)\left(\arctan(y^{1/3}) - \frac{\pi}{2} \right)+ \ln(1+y^{2/3})\right] \, ,
\ea
\be
\xi_2(\chi) = \frac{9y^2 + 7 y^{8/3}}{7\beta^2} \left[1 + y^{1/3} \left(\arctan(y^{1/3})- \frac{\pi}{2} \right)\right] \, , 
\ee
where again $y = \beta \chi$.  This can be generalized to other potentials by solving the differential equation
\be
\frac{z^2(\chi)}{z'(\chi)} = \frac{\gamma}{\beta^3} \frac{d}{d\chi}
 \left[ (\beta \chi)^{7/3} \left(\frac{9}{7} + (\beta \chi)^{2/3n} \right)^n \right] \, ,
\ee
and requiring $z$ to have the correct asymptotic behavior.  This
method again leads to an analytic potential for $n$ being an integer,
though the simplest result is that given above with $n=1$.

It is tempting to try to use functions such as the exponential and
hyperbolic tangent to parameterize $z(\chi)$.  However, we have not
found any applicable parameterization using these functions where
\emph{both} of the relevant integrals appearing in $\xi_1$ and $\xi_2$
have an analytic solution.

\section{Potentials Quadratic and Quartic in the Scalar Field}
\label{quadraticchi_section}
As mentioned in the Introduction, in order for the field $\chi$ to be dual
to the operator $\bar{q}q$ it should be a complex, matrix valued field.  So far,
we have only considered $\chi$ to be a real scalar field.  In order to address
this difficulty, one can simply promote the field with the replacement
\be
	\frac{1}{2} \chi^2 \to \chi^{\dagger}\chi,
\ee
where the extra factor of $1/2$ is introduced to give a canonical
action for a complex scalar field.  However, one will notice that all
of the potentials which were constructed in the previous section
contain fractional powers of $\chi$, and hence such a replacement is
less than desirable.  If the action is a function of $\chi^2$ only,
then such a promotion is possible.  Below, we will discuss two methods
which allow one to construct potentials which depend only on $\chi^2$

The first method we use to determine a potential which is a function of $\chi^2$ only
is to modify the ansatz (\ref{interpolatingpotential}) so that the
functions $f_1$ and $f_2$ are functions of $\phi$ only.  More
generally, let us assume

\be
	V(\phi, \chi) = \frac{e^{2\phi / \sqrt{6}}}{L^2} \left[F_0(\phi) + F_2(\phi) \chi^2 + F_4(\phi)\chi^4 \right],
\ee
with the $F$ functions to be determined.  This strategy involves
parameterizing $z(\phi)$ instead of $z(\chi)$, but the general steps
involved are similar to the cases already discussed.  The challenge is
to find a parameterization with the desired behavior that leads to an
analytic expression for $\chi(\phi)$.  The latter can be determined
from (\ref{EE2}) to be
\be
	\chi(\phi) = \pm 6^{1/4} \int_0^\phi \, 
	\left[ \frac{d}{d\phi} \ln \left( \frac{z(\phi)^2}{z'(\phi)} \right) \right]^{1/2} \,d\phi \,.
\ee
In practice, it often is easiest to start somewhere in the middle by
parameterizing some intermediate quantity such as $\chi'(\phi)$.  It
turns out that a parameterization of the form
\be
	[\chi'(\phi)]^2 = \frac{\sqrt{6}}{\phi} \left(\frac{\frac{7}{6}+\frac{3}{2}\beta \phi^n}{1+ \beta \phi^n}\right)
\ee
has the desired asymptotic behavior and leads to analytic solutions
for both $\chi(\phi)$ and $z(\phi)$.  In general, these solutions
involve hypergeometric functions.  A
notable special case is if $n= 1/2$.  Then the solutions are
\be
	z(\phi) = \frac{\gamma \phi^{1/6}}{6 \left[ (\beta \sqrt{\phi})^{1/3} - (1+\beta \sqrt{\phi})^{1/3} \right]} \, ,
\ee
\be
	\chi(\phi) = \frac{2^{3/4}}{3^{5/4} \beta}\left(G(\phi) - G(0) + \ln \left[\frac{8 + 3\sqrt{7}}{8 + 9 \beta \sqrt{\phi} + G(\phi)}\right] \right) \, ,
\ee
with
\be
	G(\phi) = 3\sqrt{(1+\beta \sqrt{\phi})(7+9 \beta \sqrt{\phi})} \,.
\ee
For this parameterization define 
\ba
	R(\phi) &=& \frac{\sqrt{6}}{2} \frac{z(\phi)}{z'(\phi)} \,,
\ea  
and make use of the calculus relations
\ba
	\phi''(z) &=& - z''(\phi) / [z'(\phi)]^3 \, ,\\
	\chi'(z) &=& \chi'(\phi) / z'(\phi) \, ,\\
	\chi''(z) &=& \frac{\chi''(\phi)}{[z'(\phi)]^2} - \frac{\chi'(\phi)z''(\phi)}{[z'(\phi)]^3} \,.
\ea 
Then the background equations (\ref{EE3}) and (\ref{EE5}) can be written
\be
	F_0(\phi) + F_2(\phi) \chi^2(\phi) + F_4(\phi)\chi^4(\phi) 
	= -12 -8 R(\phi) - R^2(\phi) + \frac{1}{3}R^2(\phi) \left[ \chi'(\phi)\right]^2 \,,
\label{EE3phi1}
\ee
\be
	2F_2(\phi) \chi(\phi) + 4 F_4 \chi^3(\phi) 
	= \frac{2}{3}\chi''(\phi) R^2(\phi) - \frac{\sqrt{6} R(\phi)\chi'(\phi)}{3} 
	\left\{5+R(\phi) -\frac{1}{3} R(\phi)\left[\chi'(\phi)\right]^2  \right\}.
\label{EE5phi1}
\ee
These equations can be solved to determine the $F$ functions, and hence the potential in terms of 
the known functions $\chi(\phi)$ and $R(\phi)$.  Notice that there are only two equations for 
the three $F$ functions; we have some freedom to choose one of the $F$ functions at will.  One possible choice
is $F_0 = -12 + 4 \sqrt{6} \phi - \frac{3}{2} \alpha \phi^2$, which is the same as the form
in the ansatz (\ref{interpolatingpotential}).  The resulting solutions for $F_2$ and $F_4$ are quite complicated.
In addition to this fact, the leading term in the small field expansion of $F_4$ is $\phi^{-5/6}$, which
is in contradiction with a well-defined conformal limit: $V(\phi \to 0, \chi \to 0) = -12/L^2$.  

On the basis of algebraic simplicity, an alternative is to
choose $F_4$ to be constant.  It is then straightforward to solve for
$F_0$ and $F_2$.  For
small values of $\phi$, these functions have the expansions
\be
	F_0(\phi) = -12+ \frac{\sqrt{6}}{\beta^2}x 
	\left[
		4+3 x^{1/6} + \mathcal{O}(x^{1/3}) 
	\right] \, ,
\ee
\be
	F_2(\phi) = -\frac{3}{2}-\frac{9}{2} x^{1/6}+ \mathcal{O}(x^{1/3}) \,,
\ee
with $x \equiv \beta^2 \phi$.
This solution appears to be consistent; however, it suffers from the
drawback that the potential is quite complicated, and that the small
field expansion of $F_0$ contains fractional powers of $\phi$ starting
with $\phi^{7/6}$.  Thus, in this solution the mass of the dilaton is
not well-defined.  This solution adheres to all of the
ingredients we set out in Sec. \ref{ingredientsection} except for
point \ref{item6}.

We now detail a second way to determine a potential which is a
function of $\chi^2$ and is thus a good candidate for a dynamical
model of chiral symmetry breaking.  As above, we will sacrifice point
\ref{item6} in the ingredients of the model by choosing a nonstandard
dilaton mass.

One may have noticed that the general potential (\ref{finalVsoln})
simplifies greatly in the special case $\alpha = 1$.  This corresponds to 
$m_\phi^2 L^2 = 5$.
\addchanges{
If one naively uses the AdS/CFT dictionary to compute the dimension
of the corresponding operator in this case, one finds $\Delta_\phi = 5$, which is 
nonrenormalizable.  
}
We are unsure whether it is simply coincidence that this choice simplifies the
potential greatly, or whether there is some physics hidden here.  
Such a choice is certainly nonstandard, though it could be acceptable
within the context of effective field theories.  
\addchanges{
However, it should be emphasized
that in light of the discussion in Sec. \ref{powerlaw_section}, any use of
the standard AdS/CFT dictionary in regards to the field $\phi$ should be 
approached with caution due to the fact that it does not reduce to a
free field near the AdS boundary.  
}

With the choice of $(m_\phi L)^2 = 5$, the potential (\ref{finalVsoln}) becomes\footnote{
  One should take care when making this simplification due to the presence of 
  $(\alpha -1)$ in the definition of the function $\xi_1$.
}
\be
	V(\phi,\chi) = \frac{e^{2\phi / \sqrt{6}}}{L^2} 
	\left\{
		-12 +4\sqrt{6}\phi -\frac{3}{2}\phi^2+\frac{\sqrt{6}}{2} \phi  I(\chi)-\frac{1}{4} I(\chi)^2-4I(\chi)+\frac{1}{2}\left[I'(\chi)\right]^2
	\right\},
\ee 
where $I(\chi)$ is defined in (\ref{I1chidef}).  Notice that only
one of the integrals appears, and all reference to the function
$\xi_2$ has disappeared in the potential.  At this point we can simply
parameterize $I(\chi)$ to fit our needs.  Any parameterization of
$I$ will do, provided that for small or large $\chi$ the function
behaves as
\be
	I(\chi) \to \frac{n}{2} \chi^2,
\ee
where $n$ is the desired power of the scalar field in this regime (i.e. $\chi(z) \sim z^n$).  For example,
if one chooses
\be
	I(\chi) = \frac{\chi^2}{2} \left(\frac{3 + (\beta \chi)^2}{1+ (\beta \chi)^2} \right),
\ee
with $\beta$ a constant, one gets the desired asymptotic behavior for zero quark mass. 
All factors of $\chi^2$ can be promoted to 
$\chi^{\dagger}\chi$ without any problem.  Another such parameterization is
\be
	I(\chi) = \frac{\chi^2}{2} + \frac{1}{\beta^2} \ln \left[1+\beta^2 \chi^2 \right] \, .
\ee
The solutions for the fields $\phi(z)$ and $\chi(z)$ could then be found
numerically from (\ref{I1chidef}) and (\ref{finalphisoln}).

\section{Conclusion}
\label{conclusion_section}
In this paper we addressed the problem of constructing models of
AdS/QCD.  Specifically, we showed how to construct a potential for
dilaton and scalar fields that leads to an AdS/QCD model with many
essential features of QCD.  Given a suitable parameterization
$z(\chi)$ or $z(\phi)$, we show how to construct a potential
$V(\phi,\chi)$ which will have a solution with the desired properties;
as such, the main results of this paper are (\ref{finalphisoln}) and
(\ref{finalVsoln}).  Linear radial trajectories, conformal symmetry
breaking, and both spontaneous and explicit chiral symmetry breaking
can be incorporated.  The desired ingredients of the model as detailed
in Sec. \ref{ingredientsection} are numerous, and the fact that such a
solution can be dynamically generated at all within such a simple
setup is somewhat surprising.  It is especially interesting that the
mass of the dilaton can be kept arbitrary throughout much of the
analysis.
\addchanges{
However, as discussed in Sec. \ref{powerlaw_section}, traditional field theory 
interpretations of dual operators via the standard AdS/CFT dictionary
are questionable, except for particular values of the dilaton mass.
}
Explicit examples of potentials were given, at least for
the case when the light quark mass is zero, although this was only for
purposes of illustration and not a limitation in principle.  We have
also not discussed the stability of the examples we provide, as that
analysis would be beyond the scope of this work.  Before calculating
physical observables within the context of such a solution, its
stability would have to be checked.  In order to incorporate chiral
symmetry breaking, the scalar field we introduce into our action needs
to be promoted to a matrix valued field, and this requirement
complicates the analysis.  However, we find that a nonstandard
dilaton mass choice greatly simplifies the results.

One might have expected to find simple expressions for the potential
since the Lagrangian for QCD is so simple to write down.  However,
there is no reason for this to be so and, unfortunately, the analytic
expressions in our illustrative examples are not so simple, although
the general structure is.  It remains a challenge to find a simple
expression for the potential that leads to the desired properties for
the dilaton and scalar fields.  Nevertheless, the potentials given
here can be useful in the context of hadronic structure.  We also
believe that the methods we have outlined here may be useful for the
determination of potentials in other bottom up AdS/QCD models,
and at finite temperature.

\section{Acknowledgments}
We would like to thank B. Batell and T. Kelley for comments on the
manuscript, and especially T. Gherghetta for insightful comments and
discussions.  This work was partially supported by the US Department
of Energy (DOE)
under grant DE-FG02-87ER40328.  TS gratefully
acknowledges a Doctoral Dissertation Fellowship from 
and the Graduate School at the University of Minnesota.

\bibliographystyle{c:/Users/Todd/Documents/Physics/AdS_QCD/Drafts/Bibliography_Files/utphys}

\phantomsection
\addcontentsline{toc}{section}{References}
%\bibliography{AdSCFT}
\bibliography{c:/Users/Todd/Documents/Physics/AdS_QCD/Drafts/Bibliography_Files/AdSCFT}

\end{document}